\documentclass[11pt]{article}
\usepackage{nyrl,palatino}
\usepackage{graphicx}
\usepackage{amsmath}
\usepackage{amssymb}
\usepackage{cleveref}

\newcommand{\meanIRR}{20}
\newcommand{\stdIRR}{2.5}

\newcommand{\meanDealSize}{50}
\newcommand{\stdDealSize}{25}
\newcommand{\logCorrelation}{-0.3}
\newcommand{\initialFunds}{500}
\newcommand{\horizonQuarters}{12}

\newcommand{\dealsPerYear}{12.0}
\newcommand{\hurdleIRR}{15}

\newcommand{\nSimulations}{1000}

\newcommand{\adpMeanIRR}{23.6}
\newcommand{\outperformance}{2.5}

\title{Optimal Capital Deployment Under Stochastic Deal Arrivals: A Continuous-Time ADP Approach}

\author{
\textbf{Kunal Menda} \\
AltLab \\
\texttt{kunal@altlab.ai}
\and
\textbf{Raphael S Benarrosh} \\
AltLab \\
\texttt{ralph@altlab.ai}
}

\begin{document}

\maketitle

\begin{abstract}
Suppose you are a fund manager with \$100 million to deploy and two years to invest it.
A deal comes across your desk that looks appealing but costs \$50 million---half of your available capital.
Should you take it, or wait for something better?
The decision hinges on the trade-off between current opportunities and uncertain future arrivals.
This work formulates the problem of capital deployment under stochastic deal arrivals as a continuous-time Markov decision process (CTMDP) and solves it numerically via an approximate dynamic programming (ADP) approach.
We model deal economics using correlated lognormal distributions for multiples on invested capital (MOIC) and deal sizes, and model arrivals as a nonhomogeneous Poisson process (NHPP).
Our approach uses quasi-Monte Carlo (QMC) sampling to efficiently approximate the continuous-time Bellman equation for the value function over a discretized capital grid.
We present an interpretable acceptance policy, illustrating how selectivity evolves over time and as capital is consumed.
We show in simulation that this policy outperforms a baseline that accepts any affordable deal exceeding a fixed hurdle rate.
\end{abstract}


\startmain % to start the main 1-2 pages of the submission.

\section{Introduction}

Capital deployment in private equity and similar asset classes is a sequential decision problem under uncertainty.
Investment opportunities arrive stochastically, vary in size and return, and must be made within a finite deployment horizon.
Deploying too early risks missing superior opportunities, while waiting too long risks underinvestment.
In practice, managers often rely on fixed hurdle rates or heuristic pacing rules, which do not adapt to changing time and capital constraints.

In the operations research and real options literature, related problems are typically framed as a decision to commit to a single investment opportunity under Poisson or uniform arrival processes~\cite{dupuis2002optimal,lange2020real}, often in discrete time and with one state variable such as project value.
These models do not address the sequential capital allocation problem faced by fund managers, where multiple heterogeneous deals arrive over time, consume capital, and compete within a finite deployment horizon.
Recent work in strategic asset allocation has considered illiquid alternative investments in fund-of-funds~\cite{luxenberg2022strategic}, but does not address stochastic deal arrivals or the fundamental trade-off between current and future opportunities.

We model the capital deployment problem as a continuous-time Markov decision process (CTMDP)~\cite{puterman2014markov} governed by the continuous-time Bellman equation~\cite{howard1960dynamic}, with arrivals following a nonhomogeneous Poisson process (NHPP)~\cite{ross2014introduction} and deal characteristics---MOIC and size---drawn from correlated lognormal distributions.
We solve it numerically using approximate dynamic programming (ADP)~\cite{powell2007approximate} on an adaptively refined time grid aligned with expected arrival increments, and quasi-Monte Carlo (QMC)~\cite{niederreiter1992random} sampling for variance reduction.
The resulting policy is interpretable to practitioners, provably optimal within the discretized model, and yields explicit acceptance thresholds that evolve with both time remaining and capital available.
We show in simulation that this policy delivers a higher portfolio return than a baseline that accepts any affordable deal exceeding a fixed hurdle rate.

\section{Problem Formulation}

We consider a fund with \textbf{initial capital} $F_0$ (e.g., \$100M), \textbf{investment horizon} $T$ (e.g., 8 quarters), \textbf{deal arrivals} governed by an NHPP with intensity $\lambda(t)$ (e.g., uniform at 12 deals/year, or seasonally varying), and \textbf{deal payoff}: MOIC $M$ and size $S$ are jointly lognormally distributed with correlation $\rho_{\log}$.

The manager's state at time $t$ is $(f, t)$, where $f$ is remaining capital.
A deal arrival presents a choice: invest or reject.
We define $M_{\text{hurdle}}$ as the minimum MOIC required to consider an investment feasible without opportunity cost considerations.

\section{CTMDP Formulation}

Let $V(f,t)$ be the \emph{expected terminal wealth above a hurdle benchmark} at time $t$ given remaining capital $f$. 
This represents the additional value created by optimal deployment decisions beyond achieving the hurdle rate of return.

The value function evolves according to the \textbf{continuous-time Bellman equation}, which balances two competing forces:

\begin{equation}
\underbrace{\frac{\partial V}{\partial t}(f,t)}_{\text{time decay}} + \underbrace{\lambda(t) \, \mathbb{E}_{S,M} \left[ \max \left\{ V(f,t), \, \underbrace{\mathbf{1}_{S \leq f}}_{\text{affordable}} \left( \underbrace{S(M-M_{\text{hurdle}})}_{\text{deal profit}} + \underbrace{V(f-S, t)}_{\text{value after investment}} \right) \right\} - V(f,t) \right]}_{\text{expected value from deal opportunities}} = 0
\end{equation}

with terminal condition $V(f,T) = 0$ (no excess value at expiry, since no further deployment decisions are possible).

The \textbf{time decay term} $\frac{\partial V}{\partial t}$ captures how the value of remaining capital decreases as the investment horizon shortens. The \textbf{deal opportunity term} $\lambda(t) \, \mathbb{E}_{S,M}[\cdot]$ captures the expected value from incoming deal opportunities, where the intensity $\lambda(t)$ determines how frequently deals arrive.
The \textbf{affordability constraint} $\mathbf{1}_{S \leq f}$ ensures we can only consider deals that fit within our remaining capital. The \textbf{maximization} $\max\{V(f,t), \mathbf{1}_{S \leq f}(S(M-M_{\text{hurdle}}) + V(f-S,t))\}$ represents the manager's choice: either reject the deal (keep value $V(f,t)$) or accept it if affordable (get immediate profit $S(M-M_{\text{hurdle}})$ plus future value $V(f-S,t)$ from remaining capital).
The \textbf{deal profit} $S(M-M_{\text{hurdle}})$ is the net return from investing size $S$ at underwritten MOIC $M$ above the hurdle rate $M_{\text{hurdle}}$ (e.g., investing \$10M at 2.0x MOIC with hurdle 1.0x yields \$10M profit). The \textbf{value after investment} $V(f-S,t)$ represents the future excess value that can be created with the remaining capital $(f-S)$.

\section{Numerical Solution via ADP}

We discretize \textbf{capital} into $N_f$ grid points and \textbf{time} into adaptive steps with expected arrivals per step $\Delta\Lambda_k \approx 0.05$, where $\Delta\Lambda_k := \lambda(t_k)\Delta t$ (or $1 - e^{-\lambda(t_k)\Delta t}$ when $\Delta\Lambda_k$ is not small).

At each grid point $f$ and time $t_k$, we work backwards from the terminal condition. First, we \textbf{sample deal characteristics} by drawing $N_{\text{QMC}}$ QMC samples $(S,M)$ from the correlated lognormal distribution. Next, we \textbf{check affordability and compute incremental value}: for each deal, verify $S \le f$, then compute
\[
\text{inc} = \mathbf{1}_{S \le f}\!\left(S(M-M_{\text{hurdle}}) + V(f-S, t_{k+1}) - V(f, t_{k+1})\right).
\]
Finally, we \textbf{update the value function}:
\[
V(f, t_k) = V(f, t_{k+1}) + \Delta\Lambda_k \, \mathbb{E}[\max(0,\text{inc})].
\]
We use Sobol sequences~\cite{sobol1967distribution} for variance reduction and linear interpolation in $f$.

\section{Experimental Validation}

To illustrate our approach, we trained an ADP policy using a fund with \$\initialFunds M initial capital and a \horizonQuarters-quarter investment horizon.
Deal arrivals follow a Poisson process with an average rate of \dealsPerYear~deals per year (assumed to be homogeneous for simplicity).
We assume that each deal returns capital five years after investment.
Calculating the internal rate of return (IRR) $r=M^{1/5}-1$, we model deal characteristics using correlated lognormal distributions: deal sizes have mean \$\meanDealSize~M and standard deviation \$\stdDealSize~M, and deal IRR has mean \meanIRR\% and standard deviation \stdIRR\%.
The correlation between log MOIC and log deal size is set to \logCorrelation. We set the hurdle IRR $r_{\text{hurdle}} = \hurdleIRR$\%.
We assume the fund manager's underwritten MOIC unbiased, and that in 95\% of cases the realized MOIC lies within a factor of 2 of the underwritten value.

\begin{figure}[h]
\centering
\begin{minipage}[t]{0.48\textwidth}
\centering
\includegraphics[width=\textwidth]{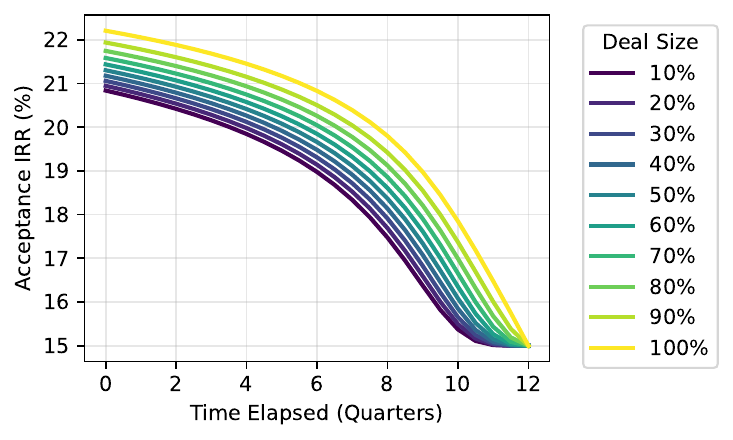}
\caption{Required IRR for different deal sizes (represented as \% of funds remaining) over time.}
\label{fig:required_irr}
\end{minipage}
\hfill
\begin{minipage}[t]{0.48\textwidth}
\centering
\includegraphics[width=0.76\textwidth]{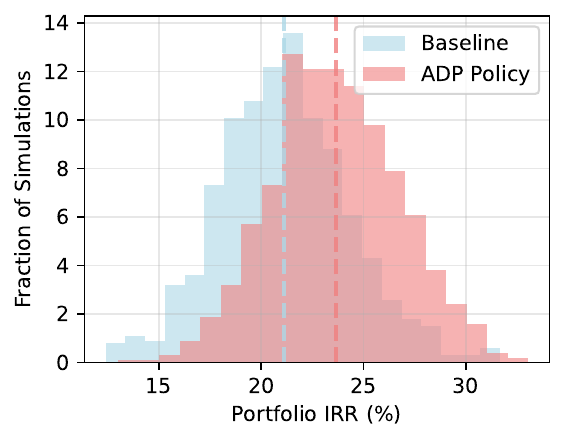}
\caption{Comparison of portfolio IRRs between baseline and ADP policy.}
\label{fig:portfolio_irr_comparison}
\end{minipage}
\end{figure}

\Cref{fig:required_irr} shows the \textbf{acceptance IRR threshold} over time for different deal sizes relative to remaining capital.
This surface reveals the optimal selectivity strategy: early in the horizon, the policy is highly selective---only deals with IRR significantly above the hurdle rate are accepted, especially for large deals that consume substantial capital.
As the horizon shortens, the threshold converges to the hurdle IRR, reflecting the manager's urgency to deploy remaining capital. Smaller deals are more easily accepted due to their lower opportunity cost and reduced impact on future flexibility.

We also benchmark the fund's performance when following the ADP-based acceptance policy shown in \Cref{fig:required_irr} against a baseline policy.
This baseline policy accepts any affordable deal with an underwritten IRR exceeding $r_{\text{hurdle}}$.
\Cref{fig:portfolio_irr_comparison} shows the \textbf{portfolio IRRs} of each policy over \nSimulations~simulations.
The ADP policy achieves a mean IRR of \adpMeanIRR\%, outperforming the baseline policy by \outperformance\%.

\section{Conclusion}
This work modeled opportunistic capital deployment as a continuous-time Markov decision process with a nonhomogeneous Poisson arrival process.
This yields an interpretable, optimal policy that adapts selectivity over time and deal size, offering a practical decision-making tool for fund managers.

\section*{Acknowledgements}
We are grateful to Dhruv Malik, Nicholas Moehle, and Anish Rai for their insights and feedback.

\bibliographystyle{unsrt}
\bibliography{bibliography}

\end{document}